\documentclass[natbib,smallextended]{svjour3_arxiv}     \smartqed  % flush right qed marks, e.g. at end of proof
\usepackage{tabularx} % extra features for tabular environment
\usepackage{amsmath}  % improve math presentation
\usepackage{amssymb}
\usepackage{graphicx} % takes care of graphic including machinery
\usepackage{subcaption}
\usepackage[normalem]{ulem}
\usepackage{dirtytalk}
\usepackage{aas_macros}
\usepackage{float}
\usepackage[colorlinks=true,citecolor=blue,urlcolor=blue,breaklinks]{hyperref}
\usepackage[utf8]{inputenc}
\usepackage[english]{babel}

\usepackage{siunitx}
\DeclareSIUnit \h {\ensuremath{\mathit{h}}}
\DeclareSIUnit \pc {pc}

\begin{document}

\title{The Cosmic Microwave Background and $H_0$}

\author{Pablo Lemos \and Paul Shah}

\institute{
P. Lemos \at 
Mila - Quebec Artificial Intelligence Institute, Montr\'{e}al, 6666 Rue Saint-Urbain, QC H2S 3H1, Canada
\and
Department of Physics, Universit\'{e} de Montr\'{e}al, Montr\'{e}al, 1375 Avenue Th\'{e}r\`{e}se-Lavoie-Roux, QC H2V 0B3, Canada
\and 
Ciela - Montreal Institute for Astrophysical Data Analysis and Machine Learning, Montréal, Canada
\and
Center for Computational Astrophysics, Flatiron Institute, 162 5th Avenue, New York, NY 10010, USA
\\
\email{pablo.lemos@umontreal.ca} \\
P. Shah \at 
  Department of Physics and Astronomy, University College London, Gower Street, London, WC1E 6BT, UK \\
\email{paul.shah.19@ucl.ac.uk}
\\ \\
Pablo Lemos: Orcid 0000-0002-4728-8473 \\
Paul Shah: Orcid 0000-0002-8000-6642
}

\maketitle

\section{The Cosmic Microwave Background}

\subsection{Introduction}

The Cosmic Microwave Background (CMB) is a fundamental component of modern cosmology, providing a wealth of information about the early universe. It is a faint radiation that permeates the entire cosmos, and it is essentially the `afterglow' of the Big Bang. The CMB is composed of photons that were released approximately $380,000$ years after the Big Bang when the universe had cooled sufficiently for atoms to form (at redshift $z \sim 1100$). Before that time, the universe was a hot, dense plasma where photons were constantly scattered by charged particles in a process known as Thomson scattering.

The CMB was first detected in 1965 by Arno Penzias and Robert Wilson at the Bell Telephone Laboratories~\citep{Penzias:1965}. They were conducting experiments using a large horn antenna designed for satellite communication.
Penzias and Wilson noticed a persistent background noise in their measurements, which they initially attributed to various sources of interference. However, they encountered difficulties in eliminating the noise, and after careful consideration, they realized that they were detecting a uniform radiation coming from all directions in the sky. They had inadvertently stumbled upon the CMB.
The discovery of the CMB provided strong support for the Big Bang theory; as opposed to the steady state theory, stating that the Universe remains unchanged~\citep{Hoyle:2000}. 

The CMB radiation is highly isotropic, with a temperature of approximately $2.7$ Kelvin. However, there is a lot of cosmological information on the tiny variations in the temperature of the CMB across the sky. These fluctuations were first detected by the Cosmic Background Explorer (COBE) satellite, launched in 1989~\citep{Smoot:1992, Smoot:1999}. 
In recent years, there have also been remarkable advancements in the detection of CMB polarization. The polarization of the CMB arises from the Thomson scattering of photons by electrons at the time of last scattering, and it carries independent information to the temperature fluctuations. The first detection of CMB polarization was announced by the Degree Angular Scale Interferometer \citep[DASI,][]{Kovac:2002} in 2002, followed by the Wilkinson Microwave Anisotropy Probe \citep[WMAP,][]{WMAP:2013}.

However, the most significant breakthroughs in CMB temperature and polarization measurements came from the Planck satellite, launched by the European Space Agency (ESA) in 2009~\citep{Planck2013}. The Planck mission provided unprecedented CMB data, unveiling the intricate anisotropy patterns of the CMB on both large and small angular scales. These observations have shed light on the conditions of the universe during the era of recombination and have provided insights into cosmological parameters, inflationary models, and the nature of dark matter and dark energy.

The origin of the modern Hubble constant tension can be reasonably dated to the first Planck release of  cosmological parameters \citep{Planck2013}, although disagreements about the value of the Hubble constant date decades back~\citep{Sandage:1958, deV}. While WMAP values for $H_0$ were compatible with other results at the time, the Planck value of $H_0 = 67.4 \pm 0.5 {\rm km \ s^{-1} \ Mpc^{-1}}$ is discrepant with the SH0ES result of $H_0 = 73.04 \pm 1.04 {\rm km \ s^{-1} \ Mpc^{-1}}$ \citep{Riess2022} at $4.9\sigma$ (although Planck is consistent with WMAP as we discuss below). As such, CMB values for $H_0$ merit considerable scrutiny. Fortunately, while local measurements focus purely on $H_0$, the CMB delivers a \say{package deal} of cosmological parameters which is highly dependent on the cosmological model. 
The cosmological model that is most supported by present-day observations is $\Lambda$CDM, consisting of a spatially-flat cosmology with a cosmological constant and three species of neutrinos with small but non-zero masses. We will make it clear when we depart from this model. As usual $h \equiv H_0 / (100$ km s$^{-1}$ Mpc$^{-1}$.
Constraints in the parameters of this model may be compared to the values from other data sets to see if other parameters are discrepant too. Also, as we explain below, the CMB is not exclusively an early universe result, as the anisotropy maps carry the imprint of late-universe physics. If therefore the Planck value for $H_0$ is considered an \say{extraordinary claim}, we discuss the evidence for it in this chapter.

\subsection{Why are there anisotropies in the CMB?}

The physical origin of density perturbations lies in the primordial universe, specifically during the period of cosmic inflation. Cosmic inflation is a theoretical concept that proposes a rapid exponential expansion of space in the early universe~\citep{Guth:1981, Linde:1982, Linde:1983}. This expansion was driven by a hypothetical scalar field known as the inflaton.
During inflation, quantum fluctuations in the inflaton field lead to variations in the energy density across space. These quantum fluctuations are inherently random and occur on extremely small scales. However, due to the exponential expansion of space during inflation, these fluctuations are stretched to cosmological scales, providing the seed for density perturbations.
After inflation, the universe enters a phase known as reheating, during which the inflaton field decays into other particles and reestablishes the thermal equilibrium of the universe. As the universe cools, the density perturbations seeded by inflation begin to grow under the influence of gravity.

The theory of inflation is widely accepted in the cosmology community, despite a lack of direct observational evidence. This is because it solves some theoretical issues, such as the horizon and flatness problems; but more importantly because of its ability to explain the origin of the density perturbations that seed the formation of cosmic structures. Inflationary models make specific predictions for the spectrum of these initial fluctuations. They are expected to be approximately scale invariant (if inflation lasts long enough), but small deviations from scale invariance can occur due to the dynamics of the inflation field. The predicted density perturbations are in excellent agreement with observations, including the temperature anisotropies in the CMB and the distribution of galaxies in the universe. Inflation, therefore, provides a natural mechanism for generating the observed structure in the universe.

The density perturbations are responsible for the observed variations in the CMB temperature across the sky. During the epoch of recombination (which occurred at the same temperature everywhere) when the universe became transparent to radiation, the photons of the CMB decoupled from the baryonic matter. At this point, the temperature of the CMB was determined by the density of matter in different regions of the universe.
Regions with higher matter density experienced stronger gravitational pull, causing photons to lose energy as they climbed out of the gravitational potential wells. As a result, these regions appear slightly cooler in the CMB. Conversely, regions with lower matter density experienced weaker gravitational pull, allowing photons to retain more energy and appear slightly warmer in the CMB. In addition, photons are Doppler-shifted by the velocity perturbations in the plasma.

During their long journey to our detectors, the CMB photons are perturbed by effects known as secondary anisotropies. In addition, foreground emission of radiation at the same frequencies as the CMB occurs both in our own galaxy and from extra-galactic sources. For the purposes of cosmology, these foregrounds must be subtracted, but we note they also carry interesting astrophysical information about the properties of dust, magnetic fields and the inter-galactic medium.

\subsection{The temperature of the CMB}

\begin{figure}[ht!]
    \centering
    \includegraphics[width=\textwidth]{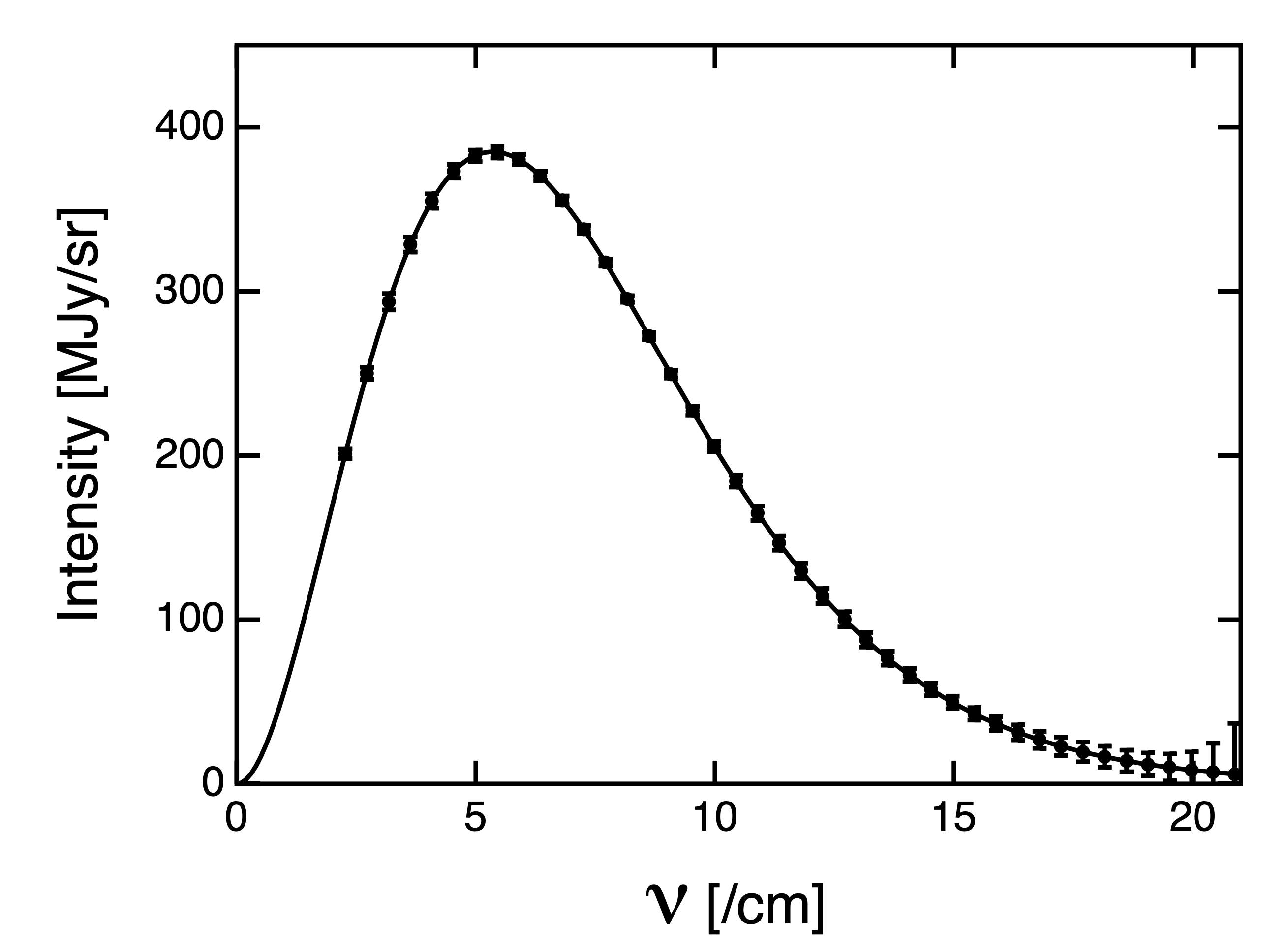}
    \caption{The spectrum of the CMB measured by FIRAS. The error bars shown are $\pm 400 \sigma$. The solid line is a $2.728 \ {\rm K}$ blackbody. Image credit:~\cite{Smoot:1997xt}}
    \label{fig:firas}
\end{figure}

The average flux intensity of the CMB observed today (that is, the monopole in the power spectrum), shown in Fig.~\ref{fig:firas} is a near-perfect black body spectrum (\textit{`the most perfect black body ever measured in nature'}, \cite{White:1999nh}), demonstrating the plasma of baryons and photons were in thermal equilibrium on the surface of last scattering (and therefore presumably at earlier times). The temperature was initially measured in balloon experiments, but was more accurately measured by the FIRAS instrument aboard the satellite COBE. The combination of data gives $T = 2.72548 \pm 0.00057$ K \citep{Fixsen2009}. Because the spectrum is thermal, the energy density of the CMB is determined solely by the temperature as $U = \frac{4}{c}\sigma T^4$. This results in the physical density of photons today as $\rho_{\gamma} = 4.65 \times 10^{-31}$ kg m$^{-3}$, or in units of the critical density $\Omega_{\gamma} \equiv \rho_{\gamma} / \rho_{c} = 5.4 \times 10^{-5}$, where the critical density represents the average mass or energy per unit volume required to achieve a flat universe.
It is worth pointing out that spectral distortions in the CMB have been proposed as a powerful cosmological probe~\citep{kogut2019cmb}, although more accurate measurements of the CMB spectrum would be required. 

We use $N_{\rm eff}$ to refer to the effective number count 
of neutrinos or other relativistic particles. Assuming there is no substantial addition to the photon density from the decay of other particles or fields, knowledge of the CMB temperature today determines the energy density of radiation at earlier times as
\begin{equation}
    \rho_{\rm r} (z) = (1+0.227 N_{\rm eff}) (1+z)^4 (\frac{T_0}{2.72548})^4 \times 4.65 \times 10^{-31} \rm kg \; m^{-3},
\end{equation}
where $T_0$ is the present-day CMB temperature.

In the standard model of particle physics, when $T \ll 1$ Mev, $N_{\rm eff} = 3.043$ (it is not an integer as the spectrum of neutrinos is not precisely a black-body; the energy density of neutrinos is altered by a contribution from the annihilation of electrons and positrons after their decoupling from the primordial plasma). However, in extensions to the standard model $N_{\rm eff}$ is often treated as a free parameter.

\subsection{The power spectrum and its features}

\begin{figure}[ht!]
    \centering
    \includegraphics[width=\textwidth]{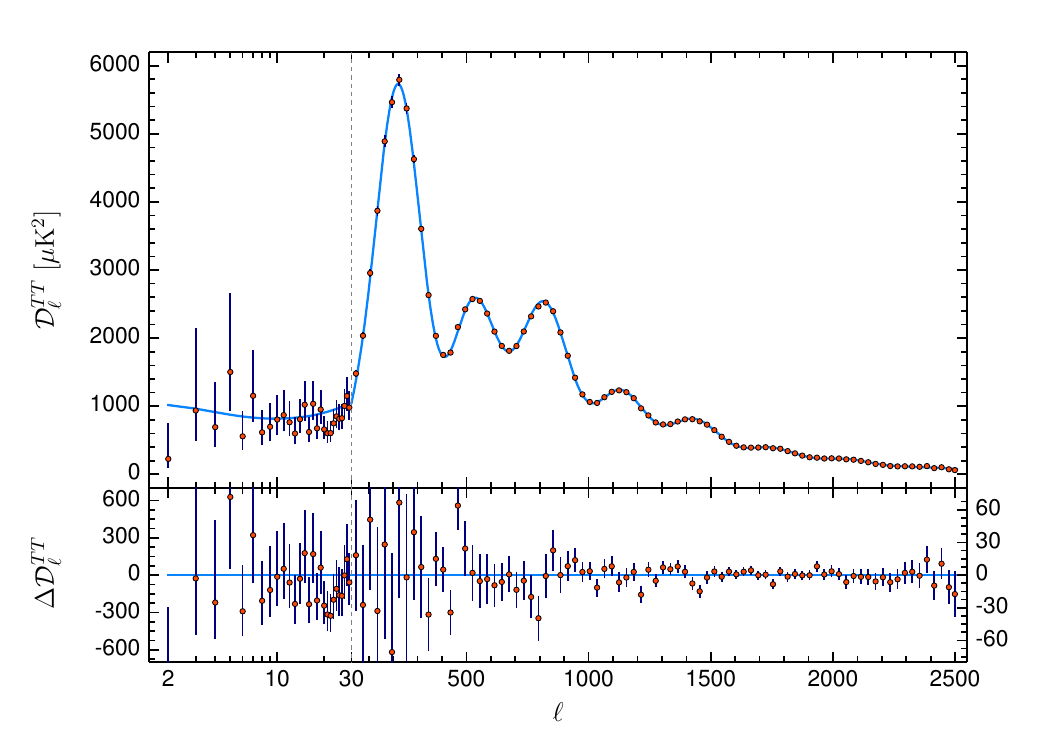}
    \caption{The binned Planck power spectrum for TT and residuals to the best fit $\mathrm{\Lambda CDM}$ model. The vertical line delineates the different methodologies used to resolve the power spectrum at $l<30$. Figure from \citet{PlanckCollaboration2018}.}
    \label{fig:PlanckTT}
\end{figure}

To analyse the anisotropies in the CMB temperature and polarization, the map of the fluctuations in the sky can be decomposed into a sum of spherical harmonics
\begin{equation}
a_{\ell m} = \int \Delta T(\theta, \phi) \; Y_{\ell m}(\theta, \phi) d\Omega \;\;,
\end{equation}
where $\ell$ is limited by the resolution of the instrument and $m$ runs over $(-\ell,\ell)$. The CMB sky represents a single sampling of the random distribution of primordial fluctuations. If we want to know the \say{true} values of cosmological parameters, we should average over a large ensemble of realisations - but we have only one! An approximation to ensemble averaging is obtained by averaging over the $2 \ell+1$ harmonics in each mode, which is likely to give a reliable answer at high-$\ell$ but is of limited accuracy at low-$\ell$, a floor referred to as cosmic variance.

The power spectra are then defined as 
\begin{equation}
\mathcal{C}^{XX}_{\ell} = \langle{|a^{XX}_{\ell m}|^2}\rangle_m \;\;,
\end{equation}
where $X$ can reperesent the temperature ($T$), E-mode polarization ($E$) or B-mode polarization ($B$). In the case of temperature, for convenience of display, the power spectra are often presented as the dimensionless quantity 
\begin{equation}
\mathcal{D}^{TT}_{\ell} = \frac{\ell(\ell+1)}{2\pi} \mathcal{C}^{TT}_{\ell}.
\end{equation} 
This scaling factor is commonly used to highlight the characteristic angular scales of the CMB fluctuations. At present, there is no evidence of non-Gaussanity in the primordial fluctuations so the $\mathcal{C}_{\ell}$ (or equivalently the 2-point correlation function) capture all available information on primary anisotropies (although there are claims of structure in low-$\ell$ $a_{\ell m}$ - see below). However, gravitational lensing induces non-zero higher-order correlation function values that can be measured separately (also see below).

\subsubsection{Primary anisotropies and length scales}

Primary anisotropies in the cosmic microwave background (CMB) radiation provide a wealth of information about the early universe and the processes that shaped its evolution. These anisotropies refer to the small fluctuations in temperature and polarization observed across the CMB sky. There are three main features of primary anisotropies in the CMB:

\begin{enumerate}
    \item {\bf The Large-Scale Plateau}: On large angular scales, typically above a few degrees, the primary anisotropies in the CMB predominantly exhibit a feature known as the large-scale plateau. This plateau corresponds to scales that were not significantly affected by pre-recombination physics and provides valuable information about the primordial Universe. 
    
    \item {\bf Acoustic Oscillations}: On smaller angular scales, below a few degrees, an additional feature becomes prominent in the primary anisotropies: the acoustic oscillations. These oscillations arise from the competition between gravitational attraction and radiation pressure in the early universe. The density and velocity perturbations in the primordial plasma give rise to sound waves that propagate through the medium, and these waves lead to periodic oscillations in the CMB power spectrum, with peaks and troughs that correspond to the specific length scale associated with the sound horizon at the time of photon decoupling. 

    The position of the peaks provides a measurement of the extrema of these oscillations at last scattering, and in particular of the {\it sound horizon at radiation drag} $r_d$, which measures the distance sounds waves can travel before recombination (\citet{White:1994}): 

    \begin{equation}
    \label{rd_intro}
    r_d = \int_{z_\star}^{\infty} \frac{c_s(z)}{H(z)} \mathrm{z}, 
    \end{equation}
    where $z_\star$ is the redshift at photon decoupling and $c_s(z)$ is the sound speed in the photon-baryon fluid (\citet{Hu:1996}) which by the standard thermodynamics of adiabatic perturbations is 
    
    \begin{equation}
    \label{cs_intro}
    c_s^2(z) = \frac{1}{3} \left[1+\frac{3}{4}\frac{\rho_b(z)}{\rho_{\gamma}(z)} \right]^{-1}. 
    \end{equation}
    The sound horizon serves as a standard ruler, and the \textit{position} of the acoustic peaks captures the history of $H(z)$ together with matter and radiation densities prior to recombination. Additionally, the \textit{heights} of the peaks represent the competition between pressure (a function of the physical baryon density $\Omega_{\rm b} h^2$) and gravity (a function of the physical matter density $\Omega_{\rm m} h^2$) as we show in Figure \ref{fig:variations} below. 

    \item{\bf Silk Damping Tail}: On even smaller angular scales, the anisotropies in the CMB exhibit a phenomenon known as the Silk damping tail. This effect arises from the diffusion of photons caused by scattering off free electrons just prior to recombination. The diffusion process tends to erase small-scale temperature fluctuations in the CMB, resulting in a damping of the power spectrum on scales below the {\it Silk scale}: 

    \begin{equation} 
    \lambda_D \sim \sqrt{N} \lambda_c \sim \sqrt{\eta_{\star} \lambda_c},
    \end{equation} 
    where $N$ is the number of steps in the random walk, $\lambda_c$ is the mean-free path to Thomson scattering, and $\eta_{\star}$ is the conformal time at recombination. The Silk scale corresponds to an angular scale around $\ell \sim 1000$ and is determined by baryon density and $H(z)$ just prior to recombination ~\citep[e.g. ][]{Hu:1997}.

\end{enumerate}

\subsubsection{Secondary anisotropies}

Secondary anisotropies in the CMB refer to additional fluctuations observed in the CMB sky, which arise from various astrophysical and cosmological phenomena. The main sources of secondary CMB anisotropies are: 

\begin{enumerate}
    \item {\bf integrated Sachs-Wolfe (ISW) effect}: The ISW effect~\citep{Sachs:1967}  is caused by the gravitational potential wells that photons traverse as they travel through evolving large-scale structures in the universe. If the potentials change photons pass through them, their energy is altered, leading to temperature fluctuations in the CMB. In particular, the Late ISW effect provides constraints on the expansion rate of the universe during the accelerated expansion era at $z<0.6$, albeit as it primarily affects large scales this constraint is limited by cosmic variance. 
    \item {\bf CMB Lensing}: The light from the CMB is deflected by foreground matter inhomogeneities, with the typical deflection angle being 3' ~\citep{lewis2006weak}. This is in the domain of weak lensing, where just one image source is seen but the source is distorted and magnified or de-magnified. The amount of lensing depends on the degree of inhomogeneity (best captured by the combination $\sigma_8 \Omega_{\rm m}^{0.25}$) and a combination of angular diameter distances between the observer, deflector and source (sometimes called the lensing efficiency). For the CMB, the lensing efficiency peaks between $1<z<2$ meaning it is sensitive to structure just beyond the reach of present-day galaxy weak lensing studies.

    Lensing has two linked effects: firstly, it smooths the observed power spectra, most noticeably for $l>800$ multipoles. Secondly, observations of the four-point correlation function of fluctuations can be used to reconstruct the (projected) foreground matter distribution. As a consistency check we may compare 1) the theory prediction of lensing given the amplitude of the fluctuations in the CMB and growth of structure 2) the observed smoothing of the power spectrum, and 3) the observed power spectrum of the lensing reconstruction. 

    \item {\bf Reionization}: During the early stages of the universe, neutral hydrogen became ionized due to the radiation from the first stars and galaxies, a process known as reionization. This reionization process leaves characteristic imprints on the CMB through the scattering and absorption of CMB photons. At $z \sim 10$ the formation of the first stars ionizes the Universe. The presence of free electrons from this point onward means that a fraction $(1-e^{-\tau})$ of CMB photons undergo Thomson scattering after reionization, where $\tau$ is the optical depth to reionization defined as the line-of-sight opacity of the CMB radiation with respect to Thomson scattering with free electrons: 

    \begin{equation} 
    \tau  = \int_0^{\chi_{\rm re}} \sigma_T \ n_e \ \mathrm{d} \chi,
    \end{equation}
    where $\chi_{\rm re}$ is the comoving distance to reionization, $\sigma_T$ is the Thomson scattering cross-section and $n_e$ is the number density of free electrons. As a consequence, the power spectrum on scales that entered the horizon before reionization is damped by a factor of $e^{-2\tau}$. 

    \item {\bf Sunyaev-Zel'dovich (SZ) effect}: The SZ effect occurs when CMB photons interact with the hot, ionized gas in galaxy clusters through inverse Compton scattering~\citep{Zeldovich:1969, Sunyaev:1972, Sunyaev:1980}. There are two types of SZ effect: thermal SZ and kinematic SZ. The thermal SZ effect arises from the Compton scattering of CMB photons with the electrons in the cluster gas, boosting the energy of the photons and distorting their spectrum to higher frequencies. The kinematic SZ effect is subdominant and arises from the bulk motion of the clusters. The thermal SZ effect can be detected by its frequency dependence, while the kinematic SZ effect is frequency independent.

\end{enumerate}

\subsection{Polarization}
\begin{figure}
    \centering
    \includegraphics[width=0.49\textwidth, angle=0]{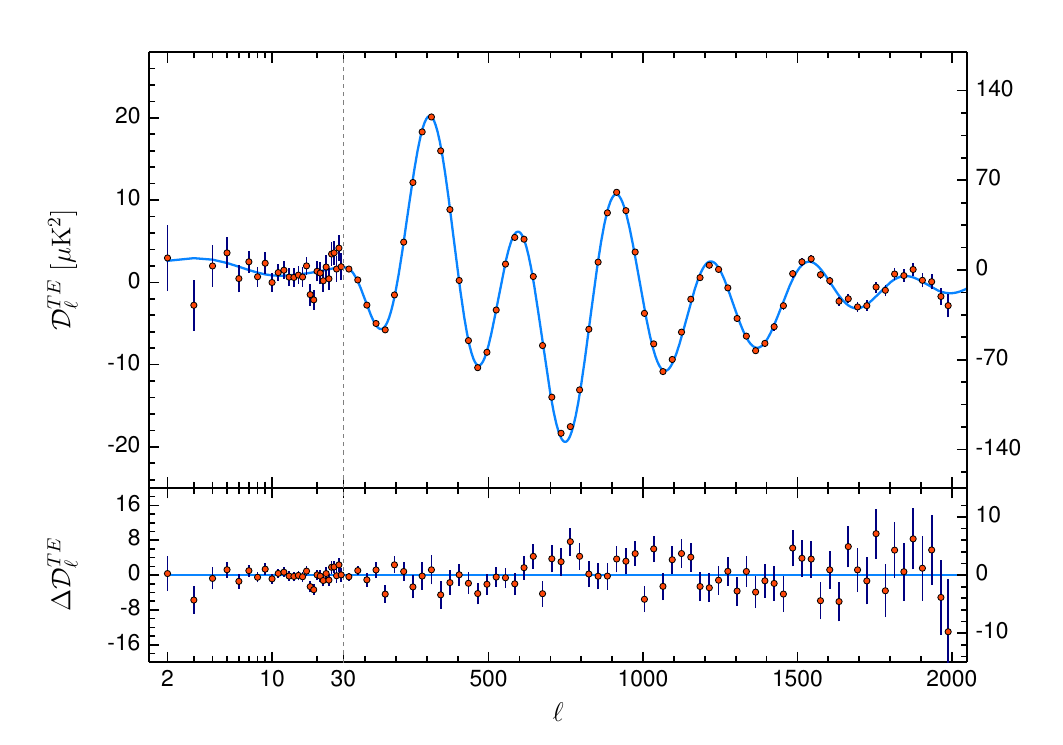}
    \includegraphics[width=0.49\textwidth, angle=0]{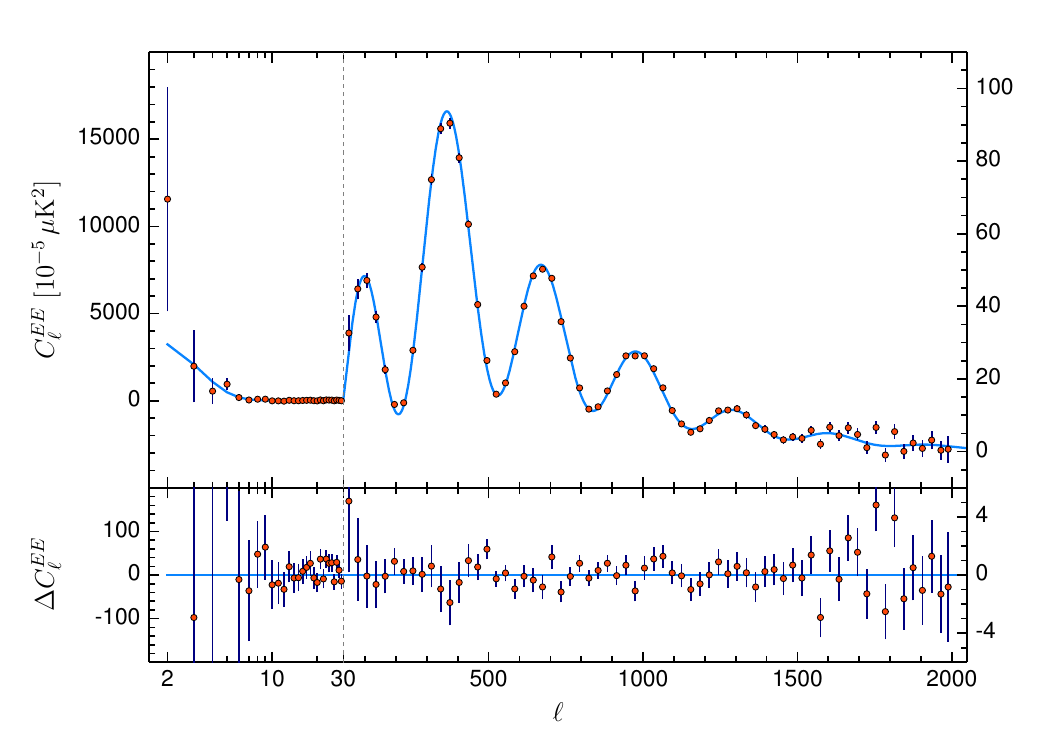} 
    \caption{{\it Planck} 2018 temperature-cross E-mode polarization power spectrum (TE, left), and E-mode autocorrelation power-spectrum (EE, right). The vertical axis on the left plot shows $\mathcal{D}_{\ell} \equiv \ell (\ell+1)C_{\ell}/(2 \pi)$. The red points are the observed data and the blue line shows the best fit ${\rm \Lambda CDM}$ cosmology. Scales change from logarithmic to linear at $\ell = 30$. The lower panels show residuals with respect to the best fits. Image credit: \cite{Planckparams18}.}
    \label{fig:planckTEEE}
\end{figure}

Temperature anisotropies are not the only source of information from CMB radiation. CMB polarization provides valuable complementary information to temperature anisotropies, offering further insights into the early Universe. The polarization pattern arises due to scalar and tensor perturbations to the metric. Scalar perturbations generate a quadrupolar component of the radiation field, leading to linear polarization through Thomson scattering during recombination. On the other hand, tensor perturbations, originating from gravitational waves, also contribute to the polarization. The polarization can be decomposed into four Stokes parameters: intensity (I), linear polarization (Q, U), and circular polarization (V). The intensity parameter represents the total power, while the linear polarization parameters capture the orientation and amplitude of the polarization. Circular polarization is absent in the CMB due to Thomson scattering properties. (for a more detailed discussion about CMB polarization see \cite{Hu:1997b, Cabella:2004}). 

The linear polarization field can be further classified into two types: E-modes and B-modes. E-modes represent the curl-free component of the polarization field and are thus primarily generated by scalar perturbations on the surface of last scattering. B-modes arise from a combination of tensor perturbations (primordial gravitational waves) and gravitational lensing of E-modes. B-modes are the curl component of the polarization field and are particularly interesting as they provide a unique window into the presence of primordial gravitational waves. Detection of B-mode polarization, after accounting for the effects of gravitational lensing, would be a direct confirmation of the existence of primordial gravitational waves, which are predicted by some inflationary models.  The {\it Planck} measurements of the E-mode power spectrum ($C^{EE}_{\ell}$) and its cross-correlation with temperature ($C^{TE}_{\ell}$) is shown in Fig.~\ref{fig:planckTEEE}. Cosmological parameters determined from the Planck polarization and temperature power spectrums are consistent with each other (see Figure 5 of \citet{PlanckCollaboration2018}).  

\subsection{Foreground removal}
Foregrounds introduced by our own Milky Way, extragalactic sources and instrument effects are primarily identified by a comparison of multiple frequency channels. The CMB has a near-perfect blackbody spectrum, and therefore a precisely defined \say{colour}. Foregrounds generated by non-thermal effects (such as synchrotron emission from dust) or thermal effects at different temperatures (such as the thermal Sunyaev-Zeldovich effect) exhibit different colours, enabling their individual identification. 

Foregrounds are removed by a combination of sky masks, frequency modelling and spectral templates $\mathcal{C}_{\ell, \rm fore}$ (for unresolved foregrounds) which represent the contribution of foreground emissions at different multipole moments. These templates capture the spectral characteristics of foreground sources such as synchrotron emission from dust and the thermal Sunyaev-Zeldovich effect. The nuisance parameters associated with these templates are marginalized over during the fitting process for cosmological parameters (see \citet{Planck2014} for a description of the process). Data from high-resolution experiments like ACT and SPT are particularly useful to constrain the nuisance parameters, and the final cleaned spectra can be compared across frequencies as a consistency check. The details are not important for this review, but the key point is: could the foreground model bias the inference of $H_0$? 

The evidence is they do not. The foreground models provide an acceptable fit across frequency ranges as measured by their $\chi^2$, justifying the claim that they correctly model the physics \citep{Planck2013}. Correlations between foreground nuisance parameters and cosmological parameters are low \citep{Planck2014}. Features such as the oscillating residuals in the region $1000<l<1500$ (see Fig. 2) and the low power for $l<30$ which play a role in driving $H_0$ for Planck lower than WMAP are present in multiple frequency channels; this suggests they are real features of the spectrum and not foregrounds. The fit to foreground-cleaned temperature spectrum is consistent with the polarisation spectrum, which depends differently on the foregrounds and in particular, is not expected to be dominated by them at high-$\ell$ \citep{Tucci2012}. Likelihoods that use different foreground subtraction methods produce cosmological parameters consistent with Planck~\citep{Dickinson:2016xyz, Sudevan:2016aqu, Wagner-Carena:2019nha}.

\subsection{The Planck Cosmology}

\begin{table}
\centering
\begin{tabular}{l l c} 
 \hline
 Parameter & Symbol & Best fit value and $68 \%$ limits \\
 \hline 
 Cold dark matter density & $\Omega_c h^2$ & $0.1202 \pm 0.0014$ \\ 
 Baryon density & $\Omega_b h^2 $ & $0.02236 \pm 0.00015$ \\
 Hubble parameter & $h $ & $0.6727  \pm 0.0060$ \\
 Optical depth to reionization & $\tau$ & $0.0544^{+0.0070}_{-0.0081}$ \\
 Scalar spectrum amplitude & $ \log \left( 10^{10} A_s \right) $ & $3.045 \pm 0.016$ \\
 Scalar spectral index & $n_s$ & $0.9649 \pm 0.0044$ \\
 \hline
\end{tabular}
\caption{Best fit values and $68 \%$ limits for the constraints from {\it Planck} 2018 TT+TE+EE+lowE \citep{Planckparams18}.}
\label{table:planck}
\end{table}

\begin{figure}[ht!]
    \centering
    \includegraphics[width=\textwidth]{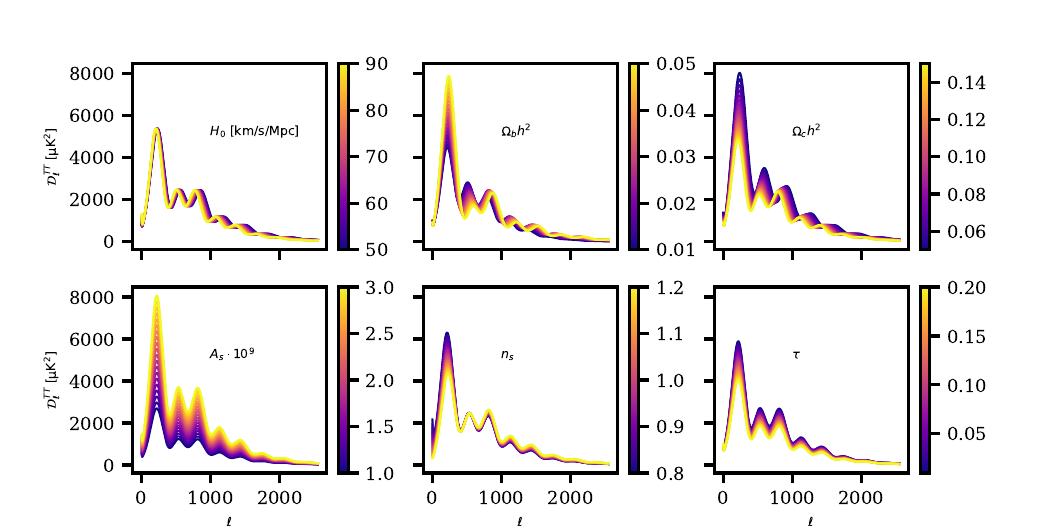}
    \caption{Variations in the CMB temperature power spectrum, as different cosmological parameters are changed, one at a time, while keeping the rest of the parameters set to the Planck best-fit values shown in Table~\ref{table:planck}.}
    \label{fig:variations}
\end{figure}

The CMB is the only cosmological probe that can simultaneously constrain all six parameters in the standard ${\rm \Lambda CDM}$ model of cosmology $\left\{ \Omega_b h^2, \Omega_c h^2, H_0, \tau, A_s, n_s \right\}$. Note that different combinations of these six parameters are commonly used to parameterize ${\rm \Lambda CDM}$. The values measured by the {\it Planck} 2018 TT+TE+EE+lowE likelihood are shown in Table~\ref{table:planck}, and the effect of varying them in the CMB power temperature power spectrum is shown in Fig.~\ref{fig:variations}.

\subsection{How does the CMB determine $H_0$?}
\subsubsection{The angular size of the sound horizon}
As we discussed in the section on primary anisotropies, the sound horizon is a characteristic length imprinted on the CMB power spectrum, visible as the positions of the multiple peaks and troughs (strictly, we should say the complete waveform of the spectrum). This is observed as a ratio $\theta_{\star} = r_s / d_A(z_{\star})$ where $d_A(z_{\star})$ is the angular diameter distance to the surface of last scattering. It is the most accurately measured parameter by Planck, and $100 \theta_{\star} = 1.0411 \pm 0.0003$ \citep{PlanckCollaboration2018}.

The theory expectation can then be simply expressed as 
\begin{equation}
\theta_{\star} = \frac{\int_{z_{\star}}^{\infty} \frac{c_s(z)}{H(z)} {\rm d}z}{\int_{0}^{z_{\star}} \frac{1}{H(z)} {\rm d}z } \;\;.
\end{equation}
At this stage, very few assumptions about cosmology have been made. The above assumes only that the sound horizon has resulted from the propagation of adiabatic pressure waves in plasma from initial seed densities, that recombination is instantaneous and photons propagate freely to us from the surface of last scattering (these last two are straightforwardly generalized).

It is convenient to define dimensionless physical densities $\omega_i = \Omega_i h^2$. In $\Lambda$CDM cosmology, we may accurately approximate the above as

\begin{equation}
\theta_{\star} = \frac{\int_{z_{\star}}^{\infty} \frac{c_s(z)}{((1+z)^3 + \frac{\omega_{\rm r}}{\omega_{\rm m}}(1+z)^4)^{1/2} } {\rm d}z}{\int_{0}^{z_{\star}} \frac{1}{((1+z)^3 +\frac{\omega_{\rm \Lambda}}{\omega_{\rm m}})^{1/2}} {\rm d}z } \;\;.
\end{equation}
The above looks rather complicated, so let us unpack it! The denominator is relatively simple, with the two terms representing the matter density and dark energy density respectively. The radiation density is mostly negligible (although it is still relevant for the first dex of expansion after recombination). 

In the numerator, we have assumed the universe is comprised only of matter and radiation. Curvature and dark energy are physically negligible pre-combination. As discussed, the radiation density $\omega_r(z, T_{\rm CMB})$ is determined by the observed CMB temperature. The sound speed $c_s(z, \omega_{\rm b})$ is determined by the physical baryon density $\omega_b$ and standard thermodynamics. The redshift of the surface of last scattering has only a weak cosmology dependence: it is approximately the freeze-out redshift when the scattering rate of photons and baryons matches the Hubble expansion rate. More baryons increase the scattering rate but also make recombination happen earlier, and the factors roughly cancel (the same argument is made for the Hubble rate). It is thus mainly a function of primordial Helium abundance $Y_P$ (which may be either measured or calculated from the baryon density, nucleosynthesis rates, and the Hubble expansion rate at that time). 

As discussed above in the section on primary anisotropies, $\Omega_{\rm m}$ and $\omega_{\rm b}$ are determined by the relative heights of the peaks of the power spectrum, and by the Silk damping scale at high-$\ell$. The only free parameter remaining to be adjusted is $h$.

\begin{figure}
    \centering
    \includegraphics[width=0.99\textwidth, angle=0]{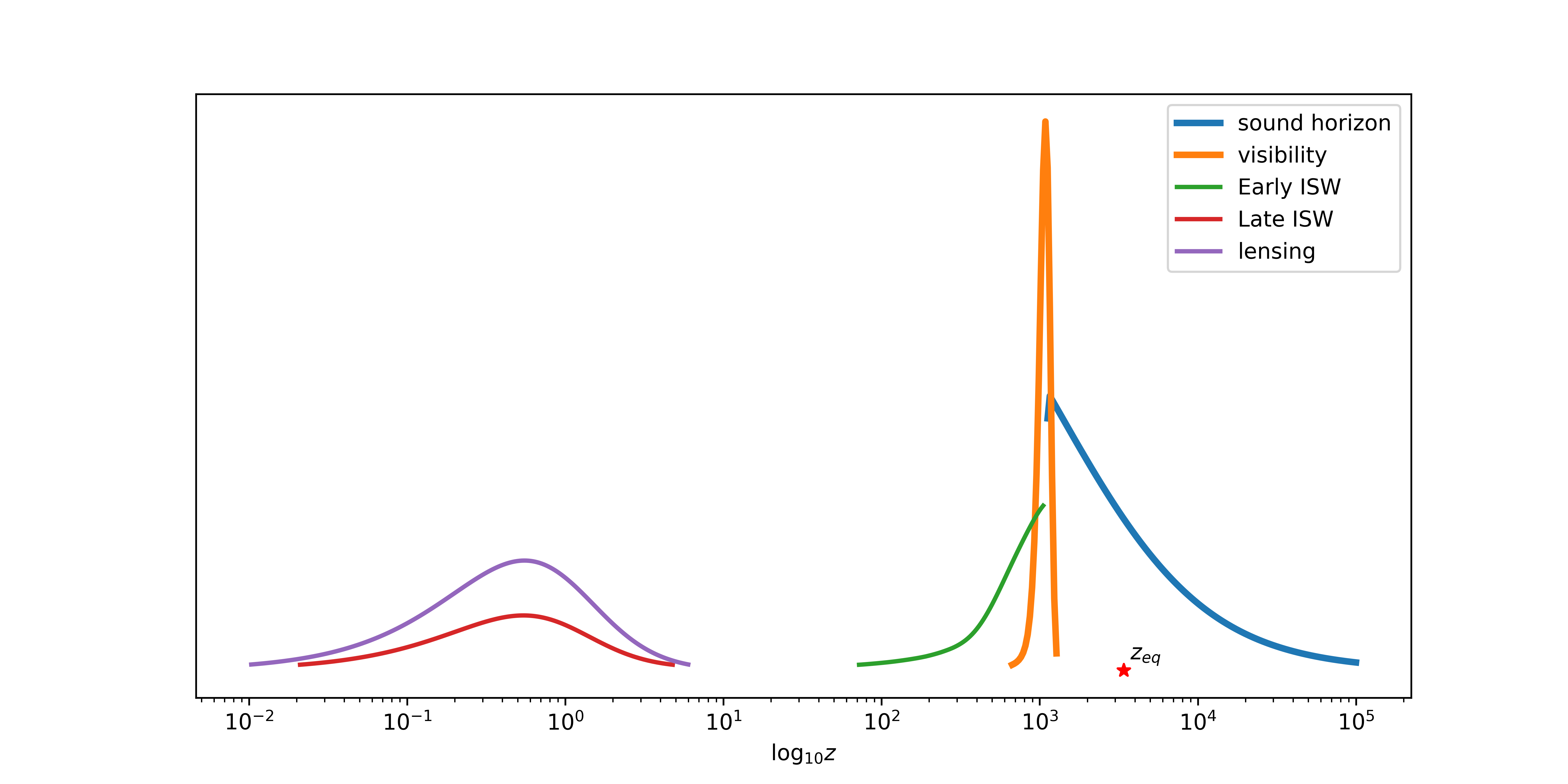}
    \caption{A schematic of the $H(z)$ redshift windows to which the CMB power spectra are sensitive to. The y-axis scale for each component is arbitrary and not shown. The growth of the sound horizon (blue) takes place primarily in the last dex of expansion before recombination : solutions to the $H_0$-tension that change the sound horizon must therefore modify $H(z)$ at this time. The visibility function (orange) represents the depth of the surface of last scattering, manifest in the Silk damping scale and polarization spectra. The Early ISW effect (green) modifies the power spectrum near its first peak due to the effect on gravitational potentials of the residual radiation density in the first dex of expansion after $z_{\rm eq}$, whereas the Late ISW effect (red) modifies the low-l power spectrum close to the onset of accelerated expansion at $z \sim 0.6$. Finally lensing (purple) peaks at a redshift corresponding to a distance that is midway (in comoving coordinates) between us and the surface of last scattering.}
    \label{fig:cmb_redshifts}
\end{figure}

The situation is a little more complicated if we drop the assumption of spatial flatness. In this case, we now have two free parameters, $\Omega_{\Lambda}$ and $h$. However, $\Omega_{\Lambda}$ is constrained by the secondary anisotropies of the CMB: the power spectrum is smoothed on a characteristic scale by gravitational lensing at $z \sim 1-2$, and power in $l<30$ modes is enhanced by the late-ISW effect (which depends on the expansion history for $z<0.6$). We show in Figure \ref{fig:cmb_redshifts} a schematic of the redshift range of the sensitivity of the CMB power spectra to $H(z)$ : each window function effectively determines a ruler whose theoretically-determined length may be compared to the observed angular size. 

\subsection{Consistency of $H_0$ derived from the CMB}
Cosmological parameters influence CMB measurements in multiple ways. It is therefore useful, as a test of $\Lambda$CDM, to check the consistency of CMB-derived parameters both internally and with non-CMB data sets.

We should however remember that we are observing one sample of a Gaussian random field, limited by cosmic variance. If we make multiple consistency tests, it would be surprising if there were \textit{not} some statistical excursions from the underlying truth.

\subsubsection{Consistency with other CMB experiments}

\begin{figure}
    \centering
    \includegraphics[width=0.99\textwidth, angle=0]{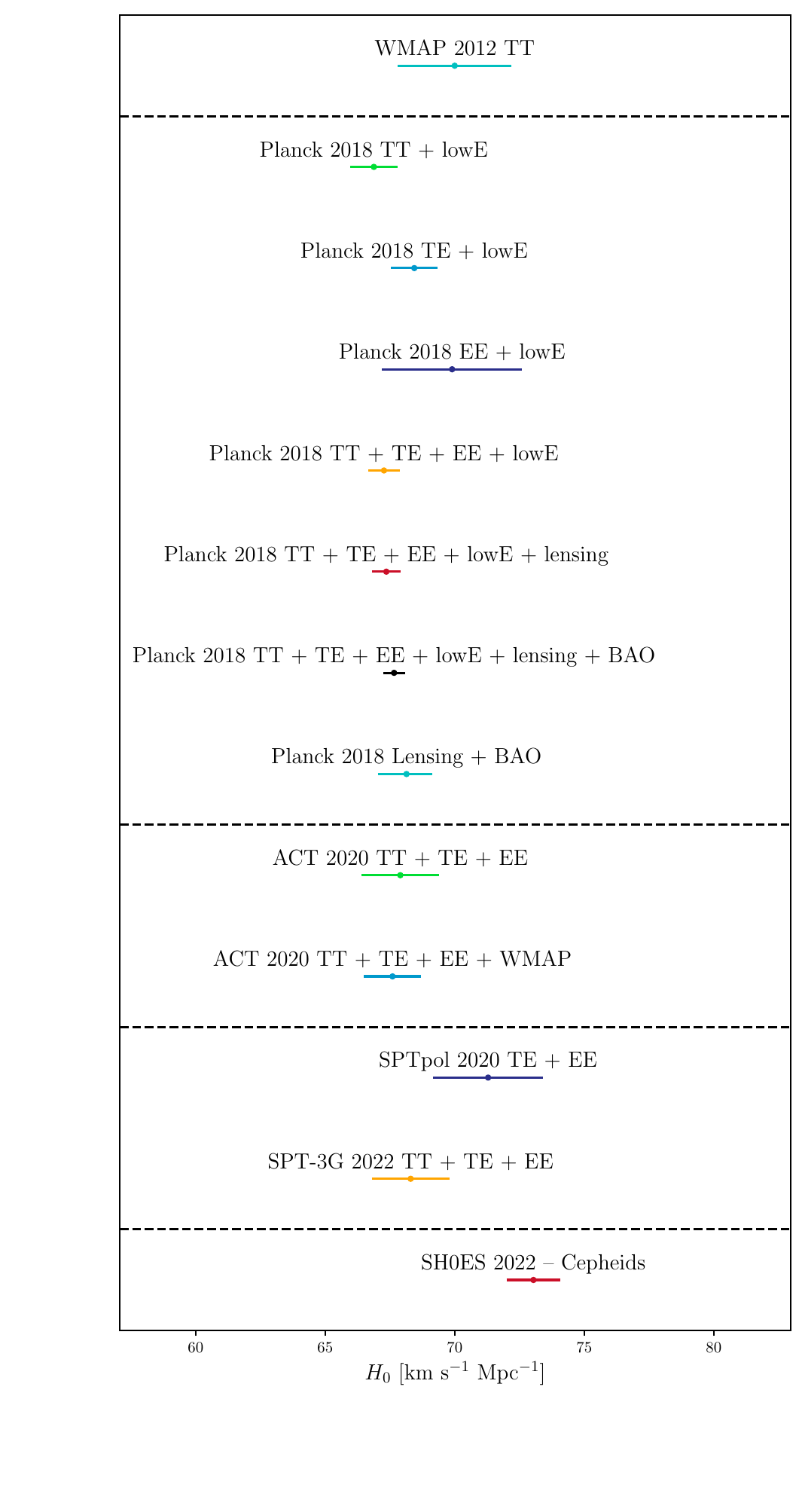}
    \caption{Comparison of Hubble constant constraints between different CMB experiments, and with the latest results from SH0ES~\citep{Riess2021}.}
    \label{fig:cmb_compare}
\end{figure}

The consistency among various cosmic microwave background (CMB) experiments provides valuable insights into the fundamental properties of the universe, particularly the Hubble constant. 
WMAP determined $H_0 = 70.0 \pm 2.2 \SI{}{\kilo\meter\per\second\per\mega\pc}$ \citep{hinshaw2013nine, Bennett2013} purely from the power spectrum of temperature fluctuations up to $l \sim 800$. This is consistent with the equivalent Planck measurement over the same multipole range (see Figure 21 of \citet{PlanckCollaboration2018}). Planck shifts $H_0$ lower due to its higher multipole range. At least part of the reason why WMAP has a higher $H_0$ than Planck appears to be due to low power in the $10<l<30$ range (as also observed by Planck - more on this below). Adding BAO information to WMAP brings the value closer to {\it Planck} ($H_0 = 68.76 \pm 0.84 \SI{}{\kilo\meter\per\second\per\mega\pc}$). A remarkable result is a consistently low value of $H_0$ from {\it Planck} lensing and BAO alone ($H_0 = 68.76 \pm 0.84 \SI{}{\kilo\meter\per\second\per\mega\pc}$ \citet{Carron:2022eyg}). 

Two notable recent experiments are the Atacama Cosmology Telescope (ACT), and the South Pole Telescope (SPT) which have played a crucial role in advancing our understanding of the CMB. ACT's combined analysis of temperature and polarization data has led to a determination of the Hubble constant, with a value of $H_0 = 67.9 \pm 1.5 \SI{}{\kilo\meter\per\second\per\mega\pc}$~\citep[e.g. ][]{Louis:2017, Aiola2020}. Interestingly, initial polarization analyses from the SPT yielded a value of $H_0 = 71.3 \pm 2.1 \SI{}{\kilo\meter\per\second\per\mega\pc}$ ~\citep{Henning2018}, in tension with {\it Planck} and ACT, and closer to the value obtained using direct measurements. However, the more recent SPT-3G measurements including temperature anisotropies recovered a low value 
$H_0 = 68.3 \pm 1.5 \SI{}{\kilo\meter\per\second\per\mega\pc}$~\citep{SPT-3G:2022hvq}, consistent with other CMB experiments. Furthermore, possible issues in the analysis of~\cite{Henning2018} were raised in~\citep{Planckparams18}, as well as in chapter 4 of~\cite{LemosPortela:2018clj}. 

Therefore, there is remarkable consistency observed between $H_0$ measurements from different CMB experiments, as shown in Fig.~\ref{fig:cmb_compare}, despite small tensions in other parameters~\citep{handley2021quantifying}. This convergence of results from independent experiments provides strong evidence for the low $H_0$ from the CMB and rules out the option of explaining the tension as a systematic effect in one of the experiments.

\subsubsection{Internal consistency of Planck data}

\begin{figure}
    \centering
    \includegraphics[width=0.99\textwidth, angle=0]{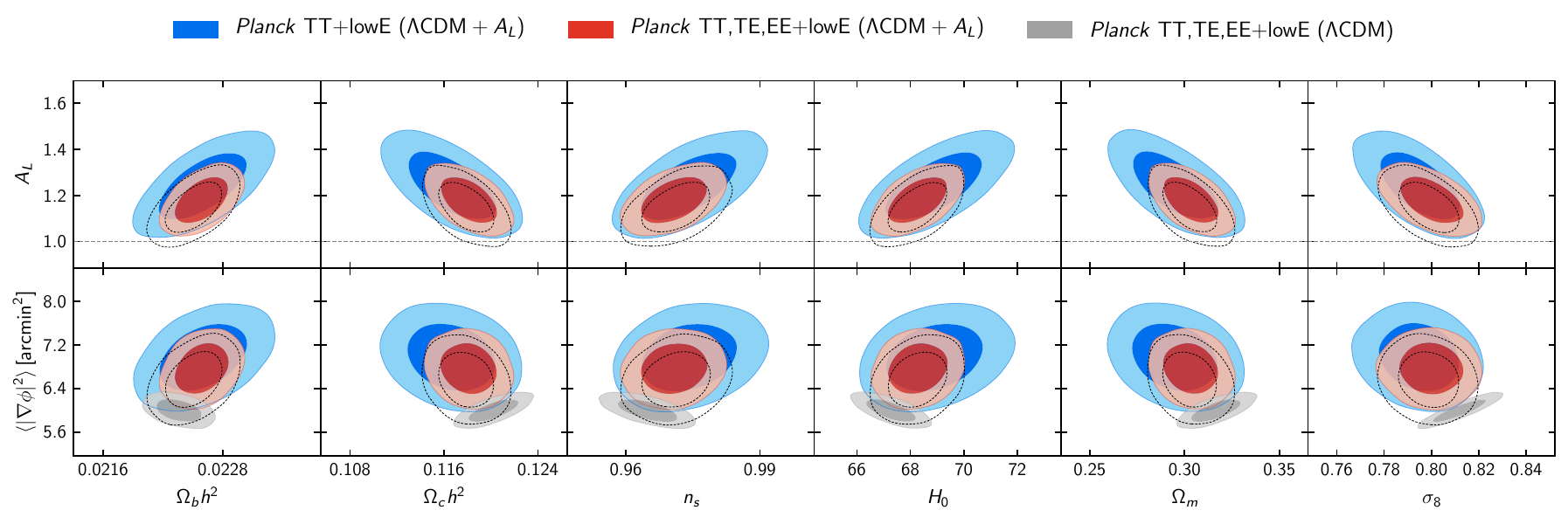}
    \caption{Comparison of the {\it Planck} 2018 $68\%$ and $95\%$ marginalized posterior distributions for different parameters, including and not including $A_L$. Note that the significance of this result decreases significantly in the CamSpec analyses \citet{Efstathiou2019}. Image credit: \cite{Planckparams18}.}
    \label{fig:planckAL}
\end{figure}

There is a moderate tension between low ($2<l<800$) and high ($801<l<2508$) multipole values for $H_0$ that has been noted by many authors (for example, see \citet{Addison2016}). The overlap between the two posteriors is consistent with the independent polarization and lensing power spectra, and also by the addition of BAO to WMAP. At least part of the difference appears to be caused by a dip in power for the multipole range $10<l<30$ (visible in Figure \ref{fig:PlanckTT} above). This seems to be a real feature of the CMB sky.

The power spectrum of lensing reconstruction and amplitude of fluctuations are consistent with each other, both internally in Planck (see Figure 3 of \citet{PlanckCollaboration2018}) and between Planck and ACT \citep{Hu2023}. However, the smoothing in the Planck power spectrum is not consistent with its amplitude, captured by the parameter $A_L = 1.180 \pm 0.065$ \citep{PlanckCollaboration2018} (where $A_L = 1$ indicates perfect consistency). The connection with $H_0$ is that allowing $A_L$ as an additional nuisance parameter in the $\Lambda$CDM solution both pulls $H_0$ to higher values and broadens the error bands, so the tension with SH0ES measurements is lowered, although it remains significant at $\sim 3.8 \sigma$ (see Fig.~\ref{fig:planckAL}). Despite its definition, $A_L$ is probably not due to lensing. It may be in part due to the dip in power between $10<l<30$: removing the $l<30$ part of the spectrum restores consistency with $A_L =1$ (and values of $H_0$ consistent with the overall Planck result). It may also be due to unknown systematics or residual foregrounds in Planck. There is no evidence for $A_L > 1$ from either ACT ($A_L = 1.01 \pm 0.11$ ~\citet{Aiola2020}) or SPT-3G ($A_L = 0.87 \pm 0.11$ ~\citet{SPT-3G:2022hvq}). In addition, the likelihood for Planck from the CamSpec analysis \citep{Efstathiou2019} including a larger area of the sky is consistent with $A_L=1$ across the full multipole range, implying it is not a real-sky effect.

Other anomalies have been noted in Planck data, for example, a lack of large angle correlations (such as would be expected from the measured power spectrum), alignment of the quadrupole and octopole, and angular variations of the spectral power and $H_0$ across patches of the sky (see for example ~\citet{Akrami:2014eta}). While some are simulated to have less than $0.1\%$ chance of occurrence in random realisations of the CMB sky, the \say{look-elsewhere} effect\footnote{We refer the reader to \url{https://en.wikipedia.org/wiki/Six_nines_in_pi} and \url{https://xkcd.com/882/} for entertaining examples.} cautions us to avoid simple interpretations of these p-values. Nevertheless, such claims are at least at the level of curiosity; but at present there is no apparent connection with the Hubble tension or any physical process.

\subsubsection{Consistency with other astrophysical data}
As we described above, the conversion of CMB spectral data into constraints on cosmological parameters is highly model dependent. Since $H_0$ from the CMB in $\Lambda$CDM is discrepant with local values, other discrepancies would add evidence to the argument $\Lambda$CDM is not the correct model and may provide hints about how to change it, especially where those parameters are correlated to $H_0$ (see Figure \ref{fig:cmb_extensions} or Figure 5 of \citet{PlanckCollaboration2018}). 

The relative heights and positions of the peaks of the CMB spectrum derive from the baryon density $\rho_{\rm b}$, the matter density $\rho_{\rm m}$ and the pre-recombination expansion history. Two key events pre-combination are nucleosynthesis (BBN) at $z_{\rm BBN} \sim 10^{10}$ and matter-radiation equality at $z_{\rm eq} \sim 3400$. Additionally, $H(z)$ at the time of recombination determines the shape of the Silk damping tail. 

The primordial abundance of helium is sensitive to the expansion rate $H(z_{\rm BBN})$ as the availability of neutrons to form it depends on the time between the decoupling of weak interactions that keep protons and neutrons in equilibrium, and the time when deuterium (D) can form. The abundance of D is very sensitive to the baryon density: as D is rapidly processed into He, a higher baryon density implies a faster interaction rate and hence less D. Primordial element abundance observations are extrapolative by nature, but the Planck values for $(N_{\rm eff}, \rho_{\rm b}) = (2.89 \pm 0.37, 0.02224 \pm 0.00022)$ (Tables 4 and 5 of P18) are consistent with observations of $Y_{\rm P} = 4n_{\rm He}/n_{\rm b} = 0.2449 \pm 0.004$ \cite{Aver2015} and $Y_{\rm D} = n_{\rm D}/n_{\rm H} \times 10^5 = 2.52 \pm 0.03$ \citep{Cooke:2018}. There is also excellent agreement with calculations using experimentally determined D-burning rates \citep{Pisanti2021}, and Planck is therefore compatible with the standard model of particle physics which predicts $N_{\rm eff} \simeq 3.043$.

It is possible to go further and drop the CMB spectrum (but not its temperature) by combining $\rho_{\rm b}$ constraints from BBN with the standard ruler provided by baryon acoustic oscillations; these are an imprint of the sound horizon on the late universe and effectively determine $\Omega_{\rm m}$ \citep{Addison2018}. In this case, one obtains $H_0 = 68.1 \pm 1.1\SI{}{\kilo\meter\per\second\per\mega\pc}$ (\citet{Cuceu2019}, see also \citet{Schoneberg2019}) and $\Omega_{\rm M} = 0.300 \pm 0.018$, fully compatible with Planck within $\Lambda$CDM.

The power spectrum $P(k)$ of matter density fluctuations has a maximum at a scale $k \sim k_{\rm eq}$ which is the scale of modes entering the horizon at $z_{\rm eq}$ (the shape near the maximum also depends weakly on $\rho_{\rm b}$ as dark matter fluctuations are suppressed by baryons). Faster expansion in the pre-recombination universe lowers the horizon and increases $k_{\rm eq}$. Therefore, the shape of the power spectrum as measured from galaxy surveys may be used to check compatibility with the Planck $H_0$ in $\Lambda$CDM. The evidence at present is in favour of compatibility with $H_0 = 69.6 \pm 1.2 \SI{}{\kilo\meter\per\second\per\mega\pc}$ (\citet{Philcox2022}, see also the chapter in this book). 

The expansion history of the universe from $z<1.2$ is strongly constrained by Type Ia SN (the furthest observations reach to $z \sim 2$). Type Ia SN do vary in intrinsic brightness by a factor of $\times 2$, but the majority of this fluctuation is attributable to the colour and duration of the explosion. The residual fluctuations may be as low as $0.07$ mag \citep{Brout2020} and the evidence points to them being a homogeneous population for the purposes of cosmology. While they cannot by themselves determine $H_0$ as this is degenerate with their fiducial absolute magnitude, the Hubble diagram constructed from the Pantheon+ sample \citep{Scolnic2022}  is fully compatible with Planck with $\Omega_{\rm m} = 0.338 \pm 0.018$ in $\Lambda$CDM \citep{Brout2022}.  Is it then is no surprise that calibrating the SN Ia magnitudes with BAO in turn produces a $H_0$ also fully compatible with Planck \citep{MacAulay2019}.

The values of $S_8 = \sigma_8 (\Omega_{\rm m}/0.3)^{0.5}$ derived from Planck (via $\Lambda$CDM assumptions on structure growth) are high (at the level of $\sim 2.5 \sigma$) compared to observations of galaxy lensing~\citep{DES:2020hen, DES:2021wwk, Heymans:2020gsg}. The relevance of this for the Hubble tension is unclear, and it has been proposed a resolution lies in correcting theory calculations of structure growth in the quasi-linear regime \citep{Amon2022}. However, we note if one were to attempt to resolve the Hubble tension by speeding up the expansion of the universe pre-recombination, one would need to compensate for the slower structure growth by adding more matter, increasing the $S_8$ discrepancy and potentially creating a new discrepancy with SN Ia. 

In summary, within the framework of $\Lambda$CDM and some simple one-parameter extensions, the only CMB-derived cosmological parameters that are inconsistent with other astrophysical datasets are $H_0$, and to a lesser extent $\sigma_8$. In particular, matter, baryon, curvature densities and number of particle species are compatible with other observations.

\subsection{The CMB in $\Lambda$CDM extensions}

\begin{figure}
    \centering
    \includegraphics[width=0.99\textwidth, angle=0]{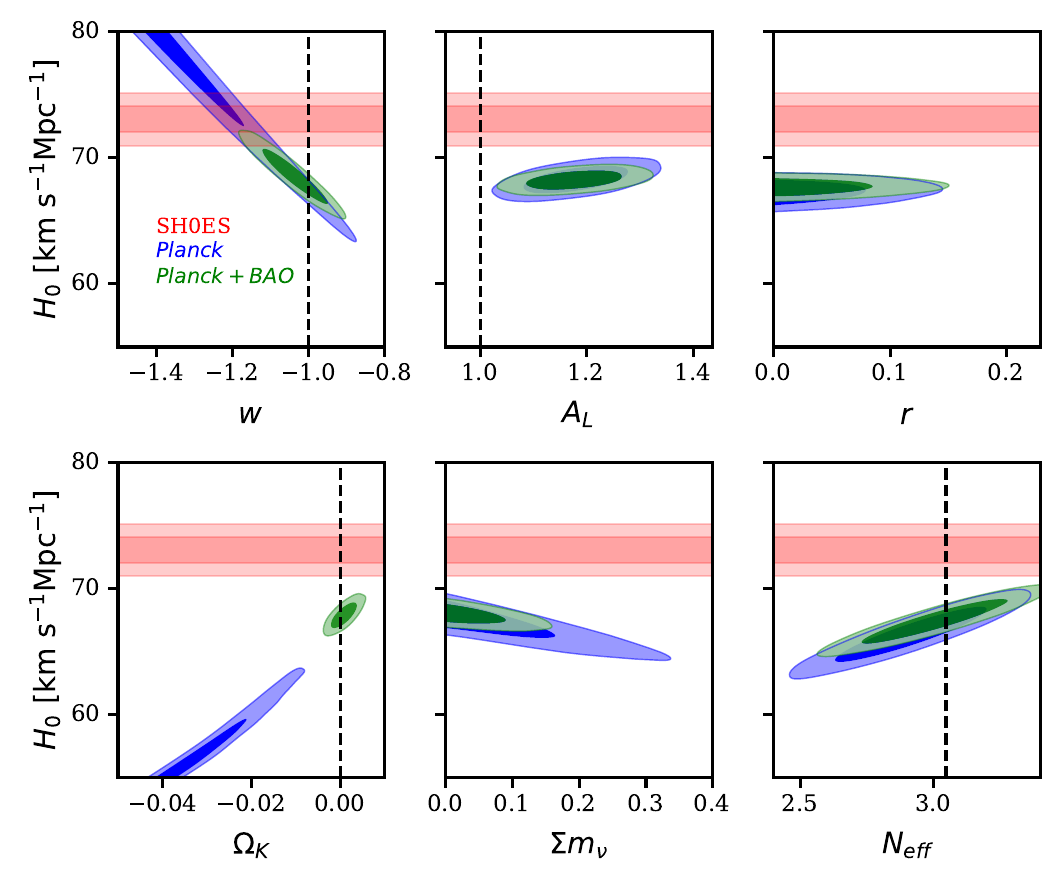}
    \caption{Comparison of Hubble constant constraints between different one parameter extensions to the $\Lambda CDM$ model, using Planck TTTEEE + lowE in blue, and adding BAO in green. In red, are the $68\%$ and $95\%$ confidence levels for $H_0$. The black, dotted lines show the $\Lambda$CDM value for each parameter.
    }
    \label{fig:cmb_extensions}
\end{figure}

Extensions to the $\Lambda$CDM model often involve modifications to the standard physics of the early universe or the presence and nature of dark energy (either pre- or post-recombination). These modifications can leave imprints on the CMB, allowing for tests of these alternative scenarios. For instance, variations in the primordial power spectrum, generated during cosmic inflation, can lead to specific patterns in the CMB temperature and polarization anisotropies. Planck data has been instrumental in placing constraints on such modifications, providing valuable insights into the inflationary paradigm and the physics of the early universe \citep{Planck:2018jri}, and finding strong support for a featureless primordial power spectrum. 

Another area of interest is the nature of dark energy itself. While the $\Lambda$CDM model assumes a cosmological constant $\Lambda$ to account for the observed accelerated expansion of the universe, alternative theories propose dynamical dark energy models or modifications to general relativity. These models can result in deviations from the standard predictions for the CMB power spectrum and other statistical properties. One common approach to parameterize dark energy deviations from $\Lambda$CDM is with a time-dependent equation of state $w(a) = w_0 + (1 - a) w_a$ ($\Lambda$CDM assumes $w_0 = -1$ and $w_a = 0$). Interestingly for the $H_0$ tension, Planck measurements are consistent with a value $w_0 < -1$, which also raises the value of $H_0$. This value falls under the regime known as "phantom" dark energy~\citep{Vikman}, as they require a second scalar field, or some other complex theoretical explanation\citep[e.g. ][]{Creminelli, Deffayet}. However, the combination of SN Ia, Planck and BAO (which may be expected to give the best constraints on dark energy) does not favour phantom dark energy \citep{Brout2022}. 

Another interesting extension to $\Lambda$CDM is spatial curvature ($\Omega_K \neq 0)$, which raised interest due to the Planck measurement $\Omega_K  = -0.044 ^{+0.018}_{-0.015}$, which is over $2 \sigma$ away from zero~\citep{Handley:2019tkm, DiValentino:2019qzk}. However, as discussed in~\cite{Efstathiou:2020wem}, this value is strongly correlated with the $A_L$ measurement from Planck discussed above, and is equally explained by statistical fluctuations in the temperature power spectra in the multipole range $800 < \ell < 1600$. Similarly to $A_L$, the CamSpec likelihood \citep{Efstathiou2019} strongly reduces the significance of this detection. Furthermore, a closed Universe leads to even lower values of $H_0$, further exacerbating the tension.  

Finally, it is possible to explore the changes to neutrino physics. For example, it is possible to vary the sum of neutrino masses $\sum m_{\nu}$ or the effective number of neutrino species $N_{eff}$ introduced above. None of these changes, however, is enough to alleviate the tension. 

A more radical idea is to increase the size of the sound horizon by introducing \say{early} dark energy (EDE) to the pre-recombination universe (possibly connected with the onset of matter domination). EDE must be present at the level of $\sim 10\%$ of the energy density during the last dex of expansion prior to the surface of last scattering, as that is when most of the growth in the size of the sound horizon happens (see Figure \ref{fig:cmb_redshifts}). Silk damping and CMB polarization on small scales constrain its presence at $z=1100$. In general, whilst EDE models cannot resolve the Hubble tension, they can alleviate it to the level of $\sim 2 \sigma$ \citep{Poulin2019}. EDE models also make specific predictions for the $l>1000$ temperature and polarization spectrums, and so future high-resolution CMB data will determine whether they are viable or not. EDE models may seem somewhat fine-tuned. They do not \textit{predict} a high late universe $H_0$ but they may \textit{explain} it (see \citet{Poulin2023} for a review). An additional problem is the boost to pre-recombination expansion requires more dark matter to reproduce the heights of the peaks of the power spectrum (as dark energy suppresses growth of structure), increasing the $S_8$ tension with galaxy weak lensing results. 

Fig.~\ref{fig:cmb_extensions} shows the effect of the most popular one-parameter $\Lambda$CDM extensions in the Planck estimates of the Hubble parameter. 
As we see, there is no unambiguous evidence from CMB data alone of a departure from Flat $\Lambda$CDM.

Generally, while it is true that one possible solution to the $H_0$ tensions is a modification to the $\Lambda CDM$ cosmological model, it is very challenging to envision new physics that solves the tension. Indeed, as described in ~\citep{Lemos:2018} to solve the tension with new physics, it needs to lower the sound horizon by approximately $9\%$ (to about $135$ Mpc) while preserving the structure of the temperature and polarization power spectra described in this chapter; and also preserving the consistency between BBN and observed abundances of light elements. 

\subsection{Conclusions and future prospects}

The cosmic microwave background (CMB) observations within the framework of the $\Lambda$CDM  model have demonstrated remarkable consistency with a wide range of non-CMB astrophysical data at varying redshifts. The cosmological parameters derived from Planck are consistent across frequencies and likelihood methodologies, and with WMAP, ACT and SPT. 

However, there are a few remaining tensions that warrant further investigation. Notably, weak lensing measurements of the parameter $\sigma_8$, which characterizes the amplitude of matter fluctuations on large scales; and, more importantly for this chapter, the local measurement of the Hubble constant ($H_0$) from the nearby universe appear to deviate from the predictions of the CMB. An interesting discrepancy that could be connected to the Hubble constant is the baseline Planck lensing amplitude parameter $A_L$. However results elsewhere are consistent with $A_L = 1$ suggesting that the observed $\sim 2 \sigma$ Planck discrepancy may be a fluke. Further investigations and independent corroborating measurements are necessary to better understand these internal inconsistencies.

Proposed solutions that aim to resolve the Hubble constant discrepancy by accelerating the growth of the sound horizon during the pre-recombination era have shown the potential to exacerbate the tension in the $\sigma_8$ parameter. These proposals, which introduce variations in the early universe's expansion rate, could have broader implications for our understanding of cosmological structures and the growth of matter fluctuations. The challenge lies in reconciling these competing tensions and finding a coherent theoretical framework that can simultaneously explain both the Hubble constant and $\sigma_8$ measurements. A comprehensive analysis of late-time evolution, combining observations from Type Ia supernovae and baryon acoustic oscillations (BAOs), offers valuable insights into the cosmic expansion history and provides constraints on different cosmological models.

Future CMB experiments hold immense promise for shedding light on unresolved questions and advancing our understanding of the universe. Next-generation missions, such as the Simons Observatory, CMB-S4, and the LiteBIRD mission (planned for launch in 2028) are poised to deliver groundbreaking observations. These experiments will allow for higher precision measurements of the CMB temperature and polarization anisotropies. This will allow tests of early dark energy models and explore potential deviations from the standard $\Lambda$CDM cosmology. The combination of these next-generation CMB experiments with independent probes, such as large-scale structure surveys and supernova observations, will provide a multi-faceted approach to addressing the outstanding tensions and advancing our understanding of the universe's evolution.

\begin{acknowledgements}
We thank Antony Lewis for comments on an earlier version of this chapter. PL acknowledges support from the Simons Foundation.
\end{acknowledgements}

\bibliographystyle{spbasic-FS-etal}      % basic style, author-year citations
\bibliography{CMB_chapter_citations, refs_pablo}% name your BibTeX data base

\end{document}